\documentstyle[epsfig,preprint,prb,aps]{revtex}

\begin{document}

\title{Resistivity due to low-symmetrical defects in metals}

\author{J. P. Dekker\footnote[1]{Present address:
Max-Planck-Institut fuer Metallforschung,
Seestr. 92, D-70174 Stuttgart, Germany}, A. Lodder and J. van Ek
\footnote[2]{Present address:
Seagate Technology,
7801 Computer Avenue South,
Bloomington, MN 55435, USA}}

\address{Faculteit Natuurkunde en Sterrenkunde, Vrije Universiteit,
         De Boelelaan 1081, 1081 HV Amsterdam, The Netherlands}

\date{\today}
\maketitle

\begin{abstract}
The impurity resistivity, also known as the residual resistivity, is
calculated {\it ab initio} using multiple-scattering theory.
The mean-free path is calculated by solving the Boltzmann equation
iteratively.
The resistivity due to low-symmetrical defects
is calculated for the FCC host metals Al and Ag
and the BCC transition metal V.
Commonly, $\frac{1}{f}$ noise is attributed to the motion of such
defects in a diffusion process.
The results for single impurities compare well to calculations by
other authors and to experimental values.
\end{abstract}


\section{Introduction}
\label{sec:introduction}

The theoretical explanation for the electrical resistivity is
well-known.
Electrons move through a regular lattice of metal atoms
without any resistance.
As soon as irregularities are introduced in this metal electrons are
scattered, which gives rise to a finite resistivity.
The temperature dependence of this quantity is mainly due to
scattering of electrons by phonons.
At zero temperature, when no phonons are present, the resistivity is
determined by defects only, such as impurity atoms.
Then it is the only remaining contribution and therefore it
is often called the residual resistivity.
In this paper the resistivity due to impurity atoms embedded in the
metal lattice is considered, the impurity resistivity, which is
extensively studied experimentally \cite{5:LanBor82}.

An interesting problem is the problem of resistance noise
\cite{5:DutHor81}.
Over a large range of frequencies the spectral density varies as
$1/f$.
This can be explained, if these resistance fluctuations arise from a kind of
diffusion process.
In most cases the frequencies range from 1 to 1000 Hz, which correspond
to typical times between jumps.
The noise is attributed to a defect, which can be of any kind, jumping
back and forth.
A simple example of such a defect is an impurity-vacancy pair, of which
we are able to calculate the resistivity for different orientations.

A lot of attempts have been made to calculate the impurity resistivity.
The simplest methods consider an atom or a cluster of atoms embedded
in free space \cite{5:BoeLodMol83,5:PapStePap94}.
More sophisticated approaches use {\it ab initio} methods like
the Korringa-Kohn-Rostoker theory (KKR)
\cite{5:MerMroZie87,5:VojMerZel92,5:EkLod91res}
to describe an impurity embedded in a metal lattice.
If this formalism is applied for two spin directions magnetic impurities
and materials can also be treated \cite{5:MerZelDed93}.
In most cases a substitutional or interstitial
\cite{5:EkLod91res,5:EkLod92res} impurity atom is considered.
In this work we mainly concentrate on the resistivity due to defects,
playing a role in substitutional electromigration, such as a vacancy,
an impurity-vacancy pair and an atom on its way to a
neighbouring vacant lattice site.
The symmetry of most of the considered defects is reduced compared to a
single impurity atom, which magnifies the required computational effort.

The theory, which is used to calculate the impurity resistivity is
described in Sec. \ref{sec:restheory}.
The theory makes use of the calculation of the electron wave function
described by Dekker {\it et al.} \cite{5:DekLodEk97FCC}, which already
requires a heavy computation of a Green's function matrix.
In Sec. \ref{sec:Alres} results are shown for the host metal Al\index{Al}.
The calculations for single $3d$
\index{Al}\index{Ca}\index{Sc}\index{Ti}\index{V}\index{Cr}\index{Mn}
\index{Fe}\index{Co}\index{Ni}\index{Cu}\index{Zn}\index{Ga}\index{Ge}
\index{As} and $4sp$ impurities in Al\index{Al} are
compared with experimental and other theoretical values in Sec.
\ref{subsec:impinAlres}.
In Sec. \ref{subsec:vacinAl} various calculations are reported,
which are interesting in view of the reliability measurements mentioned
above.
Vacancies and moving host atoms in Al\index{Al} are considered.
Resistivity calculations for impurities, a vacancy, several
impurity-vacancy pairs and and an impurity at the saddle point in
the FCC metal Ag\index{Ag} are done in Sec. \ref{sec:5spinAgres}.
Results from similar calculations for the BCC transition metal V\index{V} are
reported in Sec. \ref{sec:impinVres}.
A summary is given in Sec. \ref{sec:summaryres}.

\section{Theory}
\label{sec:restheory}

First the general theory will be presented.
After that some equations are given for the resistivity due to
low-symmetrical defects.
Finally a new expression for the generalized Friedel sum, used in the
present paper, will be given.
The conductivity of a sample can be calculated performing an
integration over the Fermi surface \cite{5:Zim72}
\begin{equation}
\label{eq:sig}
\sigma^{ij}=\frac{2 e^{2}}{{(2 \pi)}^{3} \hbar}\int_{\rm FS}
\frac{dS_{k}}{v_{k}} v^{i}_{k}\Lambda^{j}_{k},
\end{equation}
in which the velocity ${\bf v}_{k}$ of an electron with quantum numbers
$k\equiv (n{\bf k})$ is extracted from the host electronic structure.
A finite electron mean free path ${\bf \Lambda}_{k}$ is
due to the presence of defects or phonons and can be calculated by
solving the equation
\begin{equation}
\label{eq:Lambda}
{\bf \Lambda}_{k}=\tau^{0}_{k}\left[{\bf v}_{k}+
\sum_{k'}{\bf \Lambda}_{k'}P_{k'k}\right].
\end{equation}
This equation follows easily from the linearized Boltzmann equation.
In this paper scattering by a static defect is considered.
The defect can consist of a number of perturbed host atoms, an impurity
and one or two vacancies.
The probability rate $P_{k'k}$ for the transition through scattering
from state $k$ to $k'$ determines the electron lifetime
$\tau^{0}_{k}$
\begin{equation}
\label{eq:tau0}
{\tau^{0}_{k}}^{-1}=\sum_{k'}P_{k'k}.
\end{equation}
For a low concentration $c$ of a certain kind of defect,
the transition probability $P_{k'k}$ for elastic scattering is given by
\begin{equation}
P_{k'k}=2 \pi cN {\left|T_{k'k}\right|}^{2}\delta(\epsilon_{k}-\epsilon_{k'}).
\end{equation}
The calculation of the transition matrix $T_{k'k}$ requires knowledge of the
electronic wave function of the alloy.
This wave function can be calculated using multiple-scattering theory.
The formulation of this theory is given
by Dekker {\it et al.} \cite{5:DekLodEk97FCC}.
For the sake of clarity, some quantities appearing in the theory,
which are necessary in the evaluation of the impurity resistivity,
will be given here too.

The alloy wave function coefficients $c_{knL}$ and host wave function
coefficients $c^{h}_{knL}$ are related by a matrix equation,
\begin{equation}
\label{eq:Ach}
c_{knL}= \sum_{n'L'}A^{nn'}_{LL'}c^{h}_{kn'L'}.
\end{equation}
The matrix label $n$ refers to an atomic site, either at a host
position ${\bf R}_{j}$ or at an alloy position ${\bf R}_{p}$,
and $L\equiv(l,m)$ summarizes the angular momentum labels.
The matrix $A^{nn'}_{LL'}$ will be defined below.
The host wave function coefficients are evaluated at the Fermi energy
$E_{\rm F}=\kappa^{2}$ and can be written as
\begin{equation}
\label{eq:ch}
c^{h}_{knL}=-\frac
{i^{l}W^{0}_{nL}({\bf k})e^{i{\bf k}\cdot{\bf R}_{n}}}
{\sqrt{\kappa}(-{(\partial\lambda_{0}/\partial\epsilon_{k}))}^{1/2}}.
\end{equation}
The vector $W^{q}_{nL}({\bf k})$ is defined by
\begin{equation}
\label{eq:WnLk}
i^{l}W^{q}_{nL}({\bf k})=\sum_{L'}b_{LL'}({\bf k},{\bf R}_{n})
i^{l'}V^{q}_{L}({\bf k}),
\end{equation}
where $b({\bf k},{\bf R}_{n})$ is a lattice sum
\begin{equation}
\label{eq:bkn}
b({\bf k},{\bf R}_{n})=
\sum_{j'}B({\bf R}_{nj'})e^{-i{\bf k}\cdot{\bf R}_{nj'}}
\end{equation}
and $i^{l}V^{q}_{L}$ and $\lambda^{q}$ are an eigenvector and the
corresponding eigenvalue of
the KKR matrix $M({\bf k})={t^{h}}^{-1}-b({\bf k},0)$.
The matrix $B$ is defined with Gaunt coefficients $C_{LL'L''}$ and
spherical Hankel functions
$h_{L}^{+}({\bf r})=h_{l}^{+}({\kappa r})Y_{L}({\hat r})$ as
\begin{equation}
\label{eq:bnj}
B_{LL'}({\bf R})=i^{l-l'-1} \kappa \sum_{L''}C_{LL'L''}i^{l''}
h_{L}^{+}({\bf R}).
\end{equation}
It has to be stressed that the lattice sum in Eq. (\ref{eq:bkn})
extends over all host positions when ${\bf R}_{n}$ is not a host
lattice position.
When it is a host lattice position ${\bf R}_{j}$,
the corresponding term is excluded.
The $q=0$ label in Eq. (\ref{eq:ch}) refers to the eigenvalue, which
corresponds to a zero KKR matrix and thereby determines the electronic
structure of the metal.

The matrix $A^{nn'}_{LL'}$ in Eq. (\ref{eq:Ach}) is defined as
\cite{5:DekLodEk97FCC}
\begin{equation}
\label{eq:A}
A^{nn'}_{LL'}=\sum_{n_{1}L_{1}}
{{\left(1-{\cal G}^{\rm void}t\right)}^{-1}}^{nn_{1}}_{LL_{1}}
{\left(1-{\cal G}^{\rm void}t^{h}\right)}^{n_{1}n'}_{L_{1}L'},
\end{equation}
where the scattering matrices of the atomic host potentials $t^{h}_{n}$
and the ones of the atomic alloy potentials $t^{n}$ are calculated from
their phase shifts
\begin{equation}
\label{eq:t}
t^{n}_{L}=-sin(\eta_{nl})
e^{i\eta_{nl}}.
\end{equation}
The host phase shifts for an alloy position $p$, $\eta^{h}_{pl}$,
are defined to be zero, if the position $p$ does not coincide with a
host position.
The alloy phase shifts for the host position $j$ are defined to be zero
if the position does not coincide with an alloy position.

The formalism is made suitable to handle more general defects by making use
of a void system as a reference system instead of the unperturbed host.
The impurities and perturbed host atoms are replaced by free space in
this reference system.
The Green's function matrix of this reference system is calculated from
the host Green's function matrix
\begin{equation}
\label{eq:GV}
{\cal G}^{{\rm void},nn'}={\cal G}^{nn'}-
\sum_{j_{1}j_{2}}{\cal G}^{nj_{1}}
{({t^{{h}}}^{-1}+{\cal G})}^{-1}_{j_{1}j_{2}}
{\cal G}^{j_{2}n'}.
\end{equation}
The host Green's function matrix is calculated by an integration over
the Brillouin zone
\begin{equation}
\label{eq:calGnn}
{\cal G}^{nn'}= \frac{1}{\Omega_{BZ}}\int_{BZ}d^{3}k
[b({\bf k},{\bf R}_{nn'})+b({\bf k},{\bf R}_{n})M^{-1}({\bf k})
b^{{\rm T}}(-{\bf k},{\bf R}_{n'})]e^{i{\bf k}\cdot{\bf R}_{nn'}}.
\end{equation}

As derived by Van Ek {\it et al.} \cite{5:EkLod91res} and Lodder
{\it et al.} \cite{5:LodDekAth96} $T_{k'k}$ can be written within
multiple-scattering theory as
\begin{equation}
\label{eq:Tk'k}
T_{k'k}=\sum_{nL}{c_{k'nL}^{h}}^{*}T^{n}_{L}c_{knL},
\end{equation}
where $T^{n}_{L}$ is defined as
\begin{equation}
\label{eq:TnL}
T^{n}_{L}=-\frac{1}{\kappa}sin(\eta_{nl}-\eta^{h}_{nl})
e^{i(\eta_{nl}-\eta^{h}_{nl})}.
\end{equation}
We define a new quantity $Q_{knL}$ as
\begin{equation}
\label{eq:QknL}
Q_{knL}=\frac{i^{-l}}{\sqrt{\kappa}} T^{n}_{L}c_{knL}.
\end{equation}

Now the sum over $k'$ in Eq. (\ref{eq:Lambda}) can be rewritten as a
Fermi surface integral and a set of equations in terms of
$W^{0}_{nL}({\bf k})$ and $Q_{knL}$ can be derived straightforwardly.
The equation for ${\bf \Lambda}_{k}$ becomes
\begin{equation}
\label{eq:Lambda1}
{\bf \Lambda}_{k}=\tau^{0}_{k}\left[{\bf v}_{k}+
c \sum_{nn'LL'} Q_{knL}Q^{*}_{kn'L'}{\bf I}_{LL'}^{nn'}\right],
\end{equation}
where ${\bf I}$ is a Fermi surface integral with ${\bf \Lambda}_{k}$ as
a factor in the integrand
\begin{equation}
\label{eq:bfI}
{\bf I}_{LL'}^{nn'}=\frac{2 \pi}{\Omega_{\rm BZ}}\int_{\rm FS}dS_{k}
\frac{{W^{0}_{nL}({\bf k})}^{*}{\bf \Lambda}_{k}W^{0}_{n'L'}({\bf k})}
     {\left|\nabla_{k} \lambda^{0}({\bf k})\right|}
     e^{-i{\bf k}\cdot {\bf R}_{nn'}}.
\end{equation}
Eq. (\ref{eq:Lambda1}) can be solved iteratively.
In the calculation of $\tau^{0}$ we can make use of the optical 
theorem, which states that the sum over $k'$ in Eq. (\ref{eq:tau0})
can be connected to the diagonal element of the transition matrix
\begin{equation}
\label{eq:OptTh}
{\tau^{0}_{k}}^{-1}=-2c {\rm Im} T_{kk}.
\end{equation}
The comparison of the two expressions for $\tau^{0}_{k}$,
(\ref{eq:tau0}) and (\ref{eq:OptTh}), can serve as a
test for the accuracy of the Fermi surface integrals.
For a more complete description of the theory for host and alloy wavefunctions,
the reader is referred to Dekker {\it et al.} \cite{5:DekLodEk97FCC}.
Here we just add, that
an initial ${\bf \Lambda}$ has to be inserted in Eq. (\ref{eq:bfI}),
{\it e.g.} ${\bf \Lambda}_{k}=\tau^{0}_{k}{\bf v}_{k}$ or the Ziman
approximation \cite{5:Zim72}.
This leads to a new set of ${\bf \Lambda}_{k}$ according to Eq.
(\ref{eq:Lambda1}).
With this new set the integrals in Eq. (\ref{eq:bfI}) can be
recalculated.
This procedure is repeated until the new set equals the inserted set.

\begin{figure}[ht]
  \centerline{\epsfig{figure=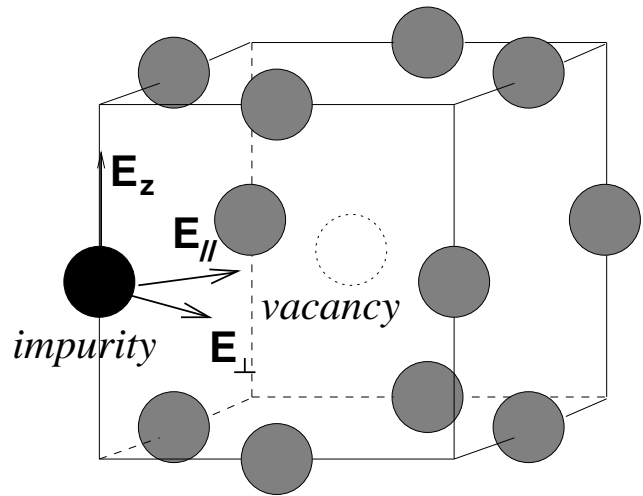,height=6.5cm}}
  \caption{Definition of the electric field directions in the FCC
           structure.}
  \label{fig:rhofcccluster}
\end{figure}

\begin{figure}[ht]
  \centerline{\epsfig{figure=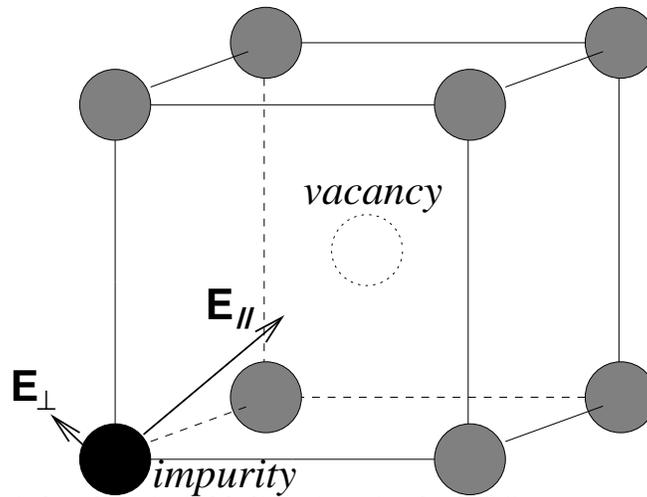,height=6.5cm}}
  \caption{Definition of the electric field directions in the BCC
           structure.
           The directions perpendicular to ${\bf E}_{\parallel}$ are
           equivalent.}
  \label{fig:rhobcccluster}
\end{figure}

Now we give the current density-field relation for a metal containing
low-symmetrical defects.
In such a metal the resistivity is anisotropic, {\it i.e.}
it depends on the direction of the current.
So, the relation between the electric field and the current density for
{\it e.g.} an impurity-vacancy pair in the FCC structure is given by
\begin{equation}
\label{eq:jFCC}
{\bf j}=\frac{1}{\rho{\parallel}}{\bf E}_{\parallel}+
\frac{1}{\rho{\perp}}{\bf E}_{\perp}+\frac{1}{\rho{z}}{\bf E}_{z},
\end{equation}
where ${\bf E}_{\parallel}$ lies along the jump direction of the
migrating atom and both
${\bf E}_{\perp}$ and ${\bf E}_{z}$ are in a perpendicular direction.
The different directions are shown in Fig. \ref{fig:rhofcccluster}.
For an impurity-vacancy pair in the BCC structure there are two
inequivalent directions, which are displayed in Fig.
\ref{fig:rhobcccluster} and therefore the current density can
be written as
\begin{equation}
\label{eq:jBCC}
{\bf j}=\frac{1}{\rho_{\parallel}}{\bf E}_{\parallel}+
\frac{1}{\rho_{\perp}}{\bf E}_{\perp}.
\end{equation}
Eqs. (\ref{eq:jFCC}) and (\ref{eq:jBCC}) describe the current density in
a sample, containing only
one kind of defect, with one particular orientation.
In a real sample the orientations of a defect are distributed randomly.
Such a distribution results in a scalar resistivity, which is given by
\begin{equation}
{\rho^{-1}_{\rm FCC}}=\frac{1}{3}\left(
\frac{1}{\rho_{\parallel}}+\frac{1}{\rho_{\perp}}+
\frac{1}{\rho_{\rm z}}
\right)
\end{equation}
for an FCC metal and by
\begin{equation}
{\rho^{-1}_{\rm BCC}}=\frac{1}{3}\left(
\frac{1}{\rho_{\parallel}}+\frac{2}{\rho_{\perp}}
\right)
\end{equation}
for a BCC metal.

Finally, in order to check the requirement of charge neutrality for the
potentials to be used, we need an expression for the generalized Friedel sum.
We will show that it is possible to derive such an expression, using
the formalism presented above.
According to Lodder and Braspenning \cite{5:LodBra94} the electron density of states
of a system $n(E)$ can be written with respect to an arbitrary reference
system as
\begin{equation}
n(E)=n^{{\rm ref}}(E)+\frac{2}{\pi}{\rm Im} \frac{d}{dE} {\rm Tr}\ {\rm ln}
\ T(E),
\label{eq:lloyd}
\end{equation}
where $T(E)$ is the $t$ matrix of the system, with respect to the reference
system.
Conventionally the unperturbed host has served as a reference
system for a dilute alloy.
For a general defect
the void system serves as the natural reference system.
In that case the $t$ matrix of the system can be written as
\begin{equation}
\label{eq:TE}
T(E)=t(1-{\cal G}^{{\rm void}}t)^{-1}.
\end{equation}
The integrated density of states $N(E_{F})=\int^{E_{F}}n(E)dE$ up
to the Fermi energy counts the total number of electrons accomodated in 
the system.
The difference in the number of electrons between the alloy and the host,
$Z_{{\rm F}}$, is found by subtracting $N^{{\rm host}}(E_{F})$
\begin{eqnarray}
\label{eq:ZF}
Z_{{\rm F}}=&&N(E_{F})-N^{{\rm host}}(E_{F})=\nonumber \\
&&\frac{2}{\pi}{\rm arg}\det t
-\frac{2}{\pi}{\rm arg}\det (1-{\cal G}^{{\rm void}}t)+\nonumber \\
&&-\frac{2}{\pi}{\rm arg}\det t^{{\rm h}}+
\frac{2}{\pi}{\rm arg}\det (1-{\cal G}^{{\rm void}}t^{{\rm h}}),
\end{eqnarray}
which is the generalized Friedel sum.
In the case of spherically symmetric scatterers this general expression
simplifies to
\begin{eqnarray}
\label{eq:ZFmt}
Z_{{\rm F}}=&&
\frac{2}{\pi} \sum_{pl} (2l+1)\eta_{l}^{p}
-\frac{2}{\pi} N^{\rm h}_{\rm cluster}\sum_{l} (2l+1)\eta_{l}^{\rm h}
\nonumber \\
&&-\frac{2}{\pi}{\rm arg}\det (1-{\cal G}^{{\rm void}}t)+
\frac{2}{\pi}{\rm arg}\det (1-{\cal G}^{{\rm void}}t^{{\rm h}}),
\end{eqnarray}
in which $N^{\rm h}_{\rm cluster}$ is the number of host atoms in the void
region.

This expression is more general than expressions used in the past,
\cite{5:LodDek94}
which only apply to simple substitutional and interstitial alloys
for which no intermediate void reference system was needed.
We will show that Eq. (\ref{eq:ZFmt}) reduces to well-established
expressions applicable to those simple systems.
In order to do this, it is useful to extend the sum in Eq. (\ref{eq:GV})
to interstitial sites.
This can be done by defining host scattering matrices for those
positions as $t^{{\rm h}}_{I}=0$.
By that the elements of the matrix
${({t^{{\rm h}}}^{-1}+{\cal G})}^{-1}=t^{{\rm h}}(1+{\cal G}t^{{\rm h}})^{-1}$
are equal to zero, when one of the two or both indices refer to an
interstitial site.
The resulting matrix equation
${\cal G}^{{\rm void}}={\cal G}-{\cal G}{({t^{{\rm h}}}^{-1}+
{\cal G})}^{-1}{\cal G}$ contains only matrices of the same dimension,
and Eq. (\ref{eq:GV}) can be rewritten as
\begin{equation}
\label{eq:Gis1minGvt}
{\cal G}={(1-{\cal G}^{{\rm void}}t^{{\rm h}})}^{-1}{\cal G}^{{\rm void}}.
\end{equation}
Note that this equation can be derived directly from Eq.
(\ref{eq:GV}) in the case of a substitutional alloy, where only
lattice sites are occupied.
In the case of an interstitial impurity the matrices are enlarged due to
the presence of the interstitial atom.

The addition of a non-scattering atom does not affect the host charge.
This can be seen from Eq. (\ref{eq:ZFmt}), and is trivial from a
physical point of view.
The matrices of the third and fourth term can be multiplied, leading to
\begin{equation}
\label{eq:GvGv}
{(1-{\cal G}^{{\rm void}}t^{{\rm h}})}^{-1}(1-{\cal G}^{{\rm void}}t)=
1-{(1-{\cal G}^{{\rm void}}t^{{\rm h}})}^{-1}{\cal G}^{{\rm void}}
(t-t^{{\rm h}})=1-{\cal G}(t-t^{{\rm h}}).
\end{equation}
Hence, the Friedel sum is given by
\begin{equation}
\label{eq:ZFsubs}
Z_{{\rm F}}=
\frac{2}{\pi} \sum_{pl} (2l+1)(\eta_{l}^{p}-\eta_{l}^{{\rm h},p})
-\frac{2}{\pi}{\rm arg}\det (1-{\cal G}(t-t^{\rm h})),
\end{equation}
which has been applied in the past to substitutional \cite{5:MolLodCol83} 
and interstitial \cite{5:OppLod87II} alloys.

\section{Impurity resistivities in Al\index{Al}}
\label{sec:Alres}

\subsection{$3d$ and $4sp$ impurities in Al
\index{Al}\index{Ca}\index{Sc}\index{Ti}\index{V}\index{Cr}\index{Mn}
\index{Fe}\index{Co}\index{Ni}\index{Cu}\index{Zn}\index{Ga}\index{Ge}
\index{As}}
\label{subsec:impinAlres}

\begin{figure}[b]
\centerline{\epsfig{figure=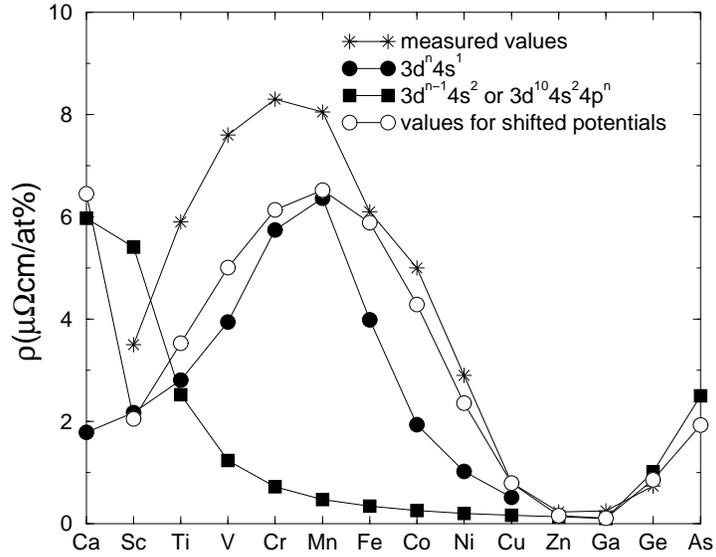,height=9.10cm}}
\caption{Impurity resistivity of $3d$ and $4sp$ atoms in Al\index{Al}. For the
$3d$ metals constructed potentials are used with either one
(filled circles) or two (filled squares) $4s$ electrons.
Results obtained for the
$4s^{2}4p^{n}$ atoms are also indicated by filled squares. Open circles
correspond to resistivity values obtained with shifted potentials for
Ca($4s^{2}$), the $3d^{n}4s^{1}$ transition metal atoms and the $4p$ atoms.
\index{Al}\index{Ca}\index{Sc}\index{Ti}\index{V}\index{Cr}\index{Mn}
\index{Fe}\index{Co}\index{Ni}\index{Cu}\index{Zn}\index{Ga}\index{Ge}
\index{As}}
\label{fig:impinAl}
\end{figure}

In this section a single $3d$\index{Sc}\index{Ti}\index{V}\index{Cr}\index{Mn}
\index{Fe}\index{Co}\index{Ni} or $4sp$\index{Cu}\index{Zn}\index{Ga}\index{Ge}
\index{As} impurity is considered embedded
in unperturbed Al\index{Al} host.
This means that the charge transfer to the surrounding host atoms
as well as lattice distortion are neglected.
Furthermore, an impurity atom has an assumed electronic configuration,
which in reality may depend on its metallic environment.
From Fig. \ref{fig:impinAl}, in which the calculated impurity
resistivities are shown, it is clear that this configuration is very
important.
The filled circles refer to calculations in which the impurity atom has
one $4s$ electron.
The values indicated by filled squares are obtained for impurity atoms
with two $4s$ electrons.
The impurity resisitivity of atoms having two $4s$ electrons decreases
with increasing atomic number, while it shows a maximum for Mn\index{Mn},
when only one $4s$ electron is present.
The experimental values \cite{5:LanBor82}, indicated in the figure by
asterisks, also show such a maximum, but the values are underestimated by
the calculations.

The potentials used in the calculations just described do not lead
to a charge neutral system, which is unphysical.
The neutrality can be restored by adding a surface charge to the
atomic spheres. \cite{5:LasSov73}
This procedure is called the shifting procedure, because it corresponds
to a shift of the atomic potential by a constant energy.
The charge of the system is calculated using the generalized Friedel sum
expression given in Sec. \ref{sec:restheory}.
This procedure has been applied to the transition metal impurities
with the ${3d}^{n}{4s}^{1}$\index{Sc}\index{Ti}\index{V}\index{Cr}\index{Mn}
\index{Fe}\index{Co}\index{Ni} electronic configuration, the
Ca\index{Ca}(${4s}^{2}$) and the
$4sp$\index{Cu}\index{Zn}\index{Ga}\index{Ge}\index{As} impurities.
The impurity resistivities, obtained with these potentials, are given
by open circles in Fig. \ref{fig:impinAl}.
The addition of charge leads to an increase of the resistivity in all
cases, except for Sc\index{Sc}, Ge\index{Ge} and As\index{As}.
The agreement with the experimental values becomes much better.
For all $4sp$\index{Cu}\index{Zn}\index{Ga}\index{Ge}
\index{As} impurities and for the transition metal impurities
with more than six $3d$ electrons the agreement is very good.

The addition of surface charge is a crude attempt to simulate the
effect of charge relaxation in the alloy.
Still, in the case of the $3d$ impurities Fe, Co and Ni it enhances the
accuracy of the resistivity significantly.
Unfortunately, this is not the case for the other $3d$ impurities.
Apparently, the surface charge does not simulate all effects of charge
relaxation in the right way.
Therefore it would be very interesting to repeat the calculations
for Sc, Ti, V, Cr and Mn with self-consistently calculated potentials.
The method of calculation of the resistivity is not affected by the
use of such potentials.

The resistivities of these impurities in Al\index{Al} were already calculated
by Boerrigter {\it et al.} \cite{5:BoeLodMol83},
Sch\"opke and Mrosan \cite{5:SchMro78} and recently by Papanikolaou
{\it et al.} \cite{5:PapStePap94}.
Sch\"opke and Mrosan \cite{5:SchMro78} used the spherical band
approximation, which means that the Fermi surface is approximated by a
sphere.
They found resistivities, which were approximately equal to the ones
following from the well-known free-electron formula of Friedel
\cite{5:Fri58}, which only contains the scattering phase shifts.
Just as the other authors mentioned they found an underestimation of
the resistivities,
which was attributed to the anisotropy of the Fermi surface.
Papanikolaou {\it et al.} \cite{5:PapStePap94} tried to
incorporate these anisotropy effects in a tricky way and
found values for the $3d$
\index{Al}\index{Ca}\index{Sc}\index{Ti}\index{V}\index{Cr}\index{Mn}
\index{Fe}\index{Co}\index{Ni}\index{Cu}\index{Zn}\index{Ga}\index{Ge}
\index{As} impurities, which were too large.
In our calculation this anisotropy is fully and consistently taken
into account, but still the impurity resistivities are underestimated.

\subsection{A migrating Al\index{Al} atom}
\label{subsec:vacinAl}

According to our calculation the resistivity of a vacancy in Al\index{Al}
is 0.57 $\mu\Omega {\rm cm/at}\%$.
We used host phase shifts for all surrounding Al\index{Al} atoms.
In first order the resistivity is the sum of the resistivities of
the separate scatterers.
Therefore it is likely that the vacancy resistivity is underestimated.
In the present case account of the scattering by the first shell
enlarges the resistivity only slightly, to 0.60
$\mu\Omega {\rm cm/at}\%$.
Our value contradicts with earlier calculations of Van Ek {\it et al.}
\cite{5:EkLod91res} who found 0.93 $\mu\Omega {\rm cm/at}\%$.

The vacancy resistivity is also extracted from simultaneous measurements
of the resistivity and the expansion of both the total volume and the
lattice constant in an Al\index{Al} sample \cite{5:SimBal60}.
In this way a value of 3.0  $\mu\Omega {\rm cm/at}\%$ is found, which is
much larger than the value we found.
This could have several reasons.
One of the reasons can be that the electronic structure of the vacancy
defect is not calculated self-consistently.
From the previous subsection indeed a strong dependence on the
electronic structure was observed.
Another reason may be that the volume expansion is not entirely due to
the absorption of vacancies or that the enlargement of the resistivity 
is not merely due to the presence of vacancies.

During a jump the resistivity changes from the initial value, via the
value at the saddle point, back to the initial value.
The saddle point value depends also on the direction of
the jump with respect to the direction of the current.
In the calculation a single saddle point atom is taken into account,
so scattering by the two small moon-shaped vacancies next to the atom
is neglected.
This procedure leads to a resistivity which is smaller than the one of
the vacancy for all directions of the current, namely
$\rho_{\parallel}=0.55 \mu\Omega {\rm cm/at}\%$ and
$\rho_{\perp}$ and $\rho_{z}$ both have the value of
$0.36 \mu\Omega {\rm cm/at}\%$.
The resistivities for the different directions are defined by Eq.
\ref{eq:jFCC}.
It is expected that the small vacancies contribute considerably to
the resistivity, leading to a value, which is larger than the
vacancy resistivity.

\begin{figure}[t]
\centerline{\epsfig{figure=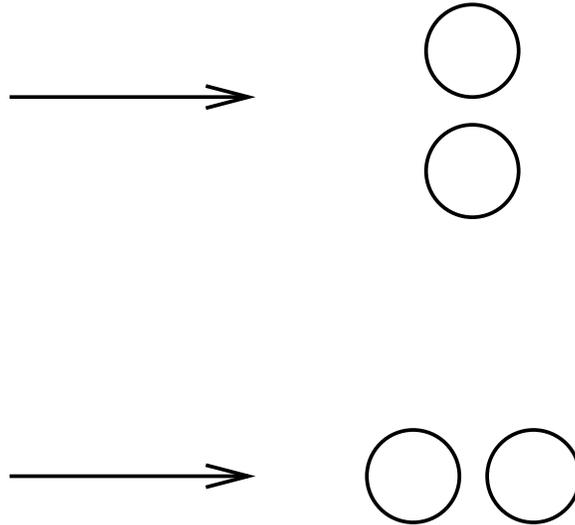,height=7.0cm}}
\caption{Vacancy pair with two different orientations with respect
to the current.
The geometrical cross-section is smaller, when the vacancy pair is
aligned with the current.}
\label{fig:twovac}
\end{figure}

Calculations for a pair of vacancies show that the resistivity,
averaged over all current
directions, is equal to the resistivity of two single vacancies.
Perhaps a larger cluster of perturbed host atoms or self-consistently
calculated phase shifts could alter this conclusion.
The symmetry of a pair is the same as the symmetry of an atom at the
saddle point.
Therefore Eq. (\ref{eq:jFCC}) holds.
The parallel resistivity $\rho_{\parallel}$ turns out to be 0.94
$\mu\Omega {\rm cm/at}\%$,
which is considerably smaller than the resistivity in the other
two directions ($\rho_{\perp}=1.24 \mu\Omega {\rm cm/at}\%$ and
$\rho_{z}=1.31 \mu\Omega {\rm cm/at}\%$).
The much smaller resistivity of a pair of vacancies
aligned along the current is easily explained intuitively
with the help of Fig. \ref{fig:twovac}.
Assuming a monotonic relation between the geometrical and scattering
cross-sections, the scattering cross-section is obviously larger
when the pair of vacancies is aligned perpendicular to the current.
However, from the results for impurity-vacancy pairs, to be presented
below, it follows that this intuitive, classical explanation does
no justice to the quantummechanical character of the scattering process.
Microscopically, one has to consider the scattering
probability due to a pair of potentials $v$ and $w$, lying at a
distance {\bf R}, which, of course, is not simply equal to the sum of the
individual probabilities too. Even in lowest order in the potential, this
probability $P_{k'k}$, calculated in
the free electron model, so using plane waves, is proportional to
\begin{equation}
\label{propfree}
P_{k'k} \sim v_{k'k}^{2} + w_{k'k}^{2}
+ 2 v_{k'k} w_{k'k} \cos ((k'-k).\bf R),
\end{equation}
in which $v_{k'k} = 4\pi \int r^{2} dr j_{0}(|k' - k|r) v(r)$
is a real quantity for a spherical potential in free space.
For a pair of vacancies $v = w$. It is clear that the cosine term
does not have a definite sign, and that the contribution will
be different for different alignments of {\bf R}. Our results for the
pair of vacancies imply, that the average contribution of this term
is positive for {\bf R} perpendicular to the current, and negative
for alignment along the current. For large
values of {\bf R} this term will average out, and the
individual probabilities just add.

\section{$5sp$ impurities in Ag
\index{Ag}\index{Pd}\index{Cd}\index{In}\index{Sn}\index{Sb}}
\label{sec:5spinAgres}

\begin{figure}[b]
\centerline{\epsfig{figure=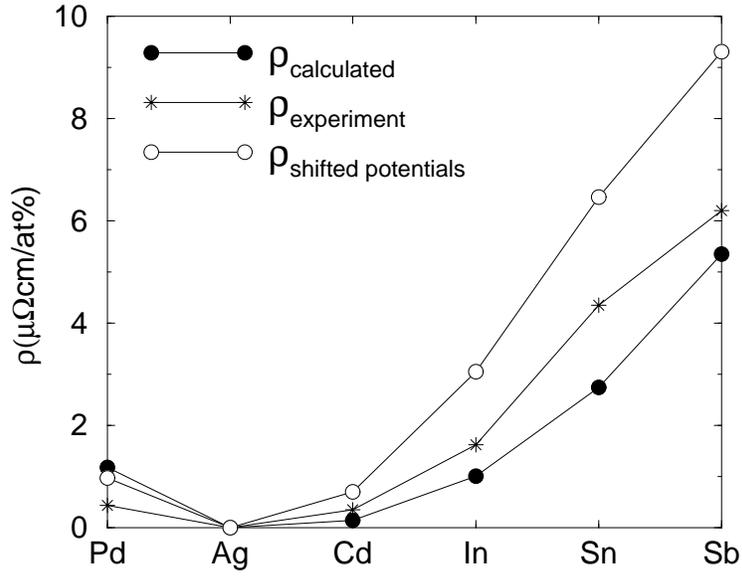,height=9.10cm}}
\caption{Calculated and measured resistivities of $5sp$ impurities
in Ag\index{Ag}.
\index{Pd}\index{Cd}\index{In}\index{Sn}\index{Sb}}
\label{fig:5spsubs}
\end{figure}

The experimentally obtained resistivities of the $5sp$ impurities
\index{Pd}\index{Cd}\index{In}\index{Sn}\index{Sb}
in Ag\index{Ag} \cite{5:LanBor82} have already been used in the analysis of
their wind valence by Dekker {\it et al.} \cite{5:DekLodEk97FCC}.
In this section the impurity resistivities will be calculated for a
single impurity, one next to a vacancy and one at the saddle point
during a diffusion jump.
In most of the calculations the perturbation of the surrounding
host atoms is not taken into account.
In Fig. \ref{fig:5spsubs} it is seen that the calculations,
indicated by filled circles,
and the measurements, indicated by asterisks,
show the same trend.
However the measured values are larger.
Only the value of 1.18 $\mu\Omega {\rm cm/at}\%$ for the $4d^{10}$
impurity Pd\index{Pd} is an overestimation.
A much lower value of 0.02 $\mu\Omega {\rm cm/at}\%$ is found,
when a $4d^{9}5s^{1}$ electronic configuration is used for the Pd\index{Pd} atom.
The experimental value of 0.44 $\mu\Omega {\rm cm/at}\%$ lies between
the two theoretical values, which suggests
that the electronic configuration is a mixture of both.
The calculated resistivities are only slightly affected by taking
into account a shell of perturbed host atoms.
A maximum increase of 0.04 $\mu\Omega {\rm cm/at}\%$ is found for In.

The shifting procedure to achieve charge neutrality is also applied in
this case.
Missing charge had to be added to the impurity.
The resulting values are indicated by open circles
in Fig. \ref{fig:5spsubs}.
Just like in the case of impurities in Al\index{Al} the resistivities are
enlarged.
However, the agreement with experiment does not improve in this
case, because the enlargement is too strong.

Similar calculations have been done by Vojta {\it et al.}
\cite{5:VojMerZel92} using self-consistent single-site potentials.
Their results are comparable to ours, but they agree somewhat better with
the experimental values.
This could be the result of a larger muffin-tin radius, they used.
Our muffin-tin radius is bounded, because of the decreased space at the
saddle point.
Nevertheless our values are reasonable.

\begin{figure}[t]
\centerline{\epsfig{figure=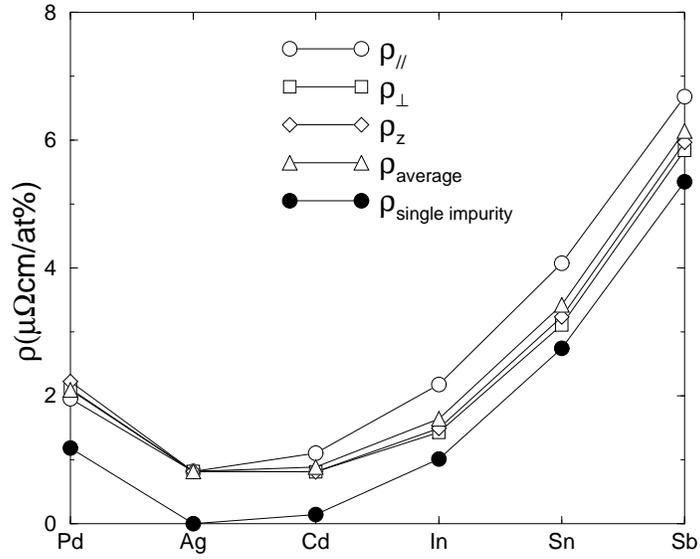,height=9.10cm}}
\caption{Calculated resistivities of pairs of a $5sp$ impurity and a
vacancy in Ag\index{Ag}. The resistivities of the single impurities are
given for comparison.
\index{Pd}\index{Cd}\index{In}\index{Sn}\index{Sb}}
\label{fig:5spvac}
\end{figure}

The resistivities for $5sp$\index{Pd}\index{Cd}\index{In}\index{Sn}\index{Sb}
impurity-vacancy pairs are given in Fig. \ref{fig:5spvac}.
The resistivity for a single vacancy is $0.82 \mu\Omega {\rm cm/at}\%$,
which is the value for Ag\index{Ag} in the figure.
The resistivity of an impurity-vacancy pair, being aligned with the
current, $\rho_{\parallel}$, is larger than the resistivity, when they
are aligned perpendicular to the current, $\rho_{\perp}$ and $\rho_{z}$.
This is in contradiction with the intuitive explanation for the
resistivity of a vacancy pair in Al\index{Al} in the different directions
in terms of a geometrical cross-section,
which is given in Sec. \ref{subsec:vacinAl} and illustrated in 
Fig. \ref{fig:twovac}. However, this behavior can be understood from the
simple expression (\ref{propfree}).
The impurity potential $w$ is certainly attractive, which corresponds
to an overall negative sign, and a vacancy potential $v$ is repulsive.
So, on the average, the cosine term in Eq. (\ref{propfree})
has the opposite sign compared with the scattering
by two vacancies. This implies a conversion of the behavior, in
agreement with or finding for the impurity-vacancy pair.
Notice also, that the resistivity of an
impurity-vacancy pair, averaged over all current directions,
$\rho_{\rm average}$, does not equal the sum of the separate
resistivities of vacancy and impurity.
The latter sum rather equals $\rho_{\parallel}$.

\begin{figure}[t]
\centerline{\epsfig{figure=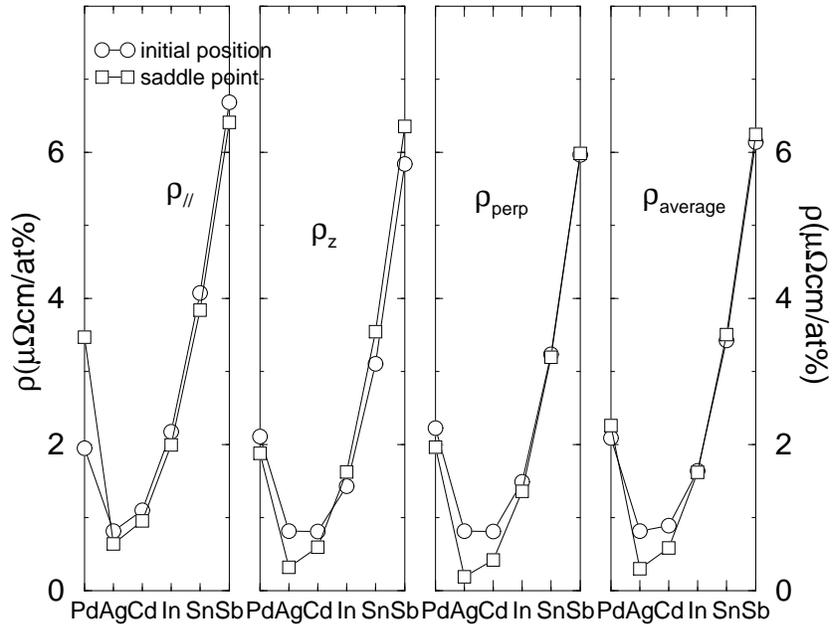,height=9.10cm}}
\caption{Calculated resistivities in Ag\index{Ag} of $5sp$ impurities, located
next to a vacancy (initial position) and at the saddle point position.
\index{Pd}\index{Cd}\index{In}\index{Sn}\index{Sb}}
\label{fig:5spzpt}
\end{figure}

In Fig. \ref{fig:5spzpt} the impurity resistivities at the saddle point
for the different current directions are compared with the corresponding
resistivities for the impurity-vacancy pair.
The saddle point resistivity follows roughly the one at the initial
position.
Again $\rho_{\parallel}$ is the largest, but for an atom at the saddle
point the cross-section is not expected to depend strongly
on the direction,
because the current sees one scattering atom from all directions.
Just like in the case of Al\index{Al}, the two small moon-shaped vacancies
around the saddle point atom are not taken into account, which is
expected to lead to an underestimation of the resistivity.

\section{Transition metal impurities in V\index{V}}
\label{sec:impinVres}

\begin{figure}[t]
\centerline{\epsfig{figure=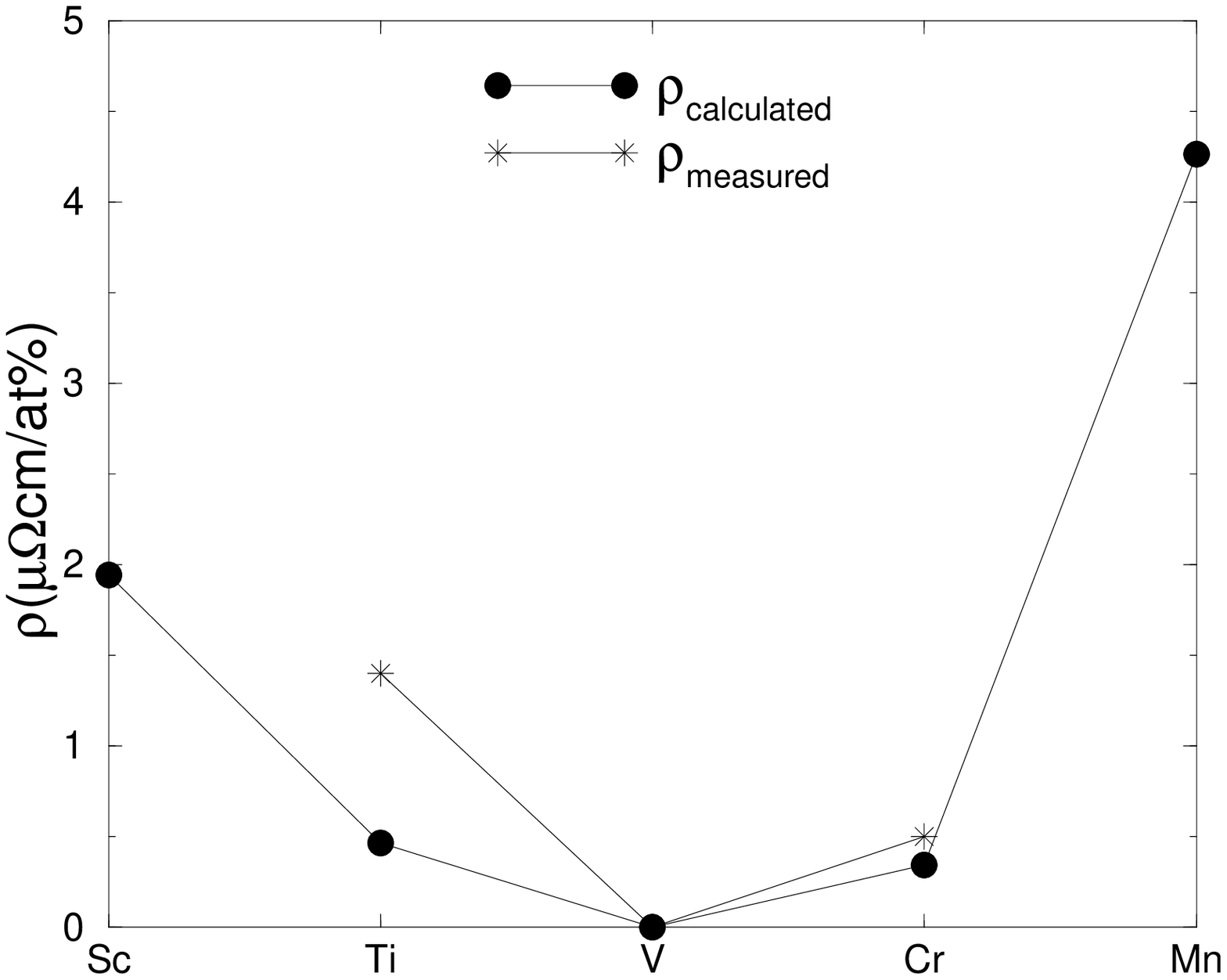,height=9.10cm}}
\caption{Calculated and measured resistivities of $3d$ impurities in
V\index{V}.\index{Sc}\index{Ti}\index{Cr}\index{Mn}}
\label{fig:3dsubsV}
\end{figure}

The measured resistivities of the $3d$ impurities Ti\index{Ti} and
Cr\index{Cr} \cite{5:LanBor82} and the calculated ones of Sc\index{Sc},
Ti\index{Ti}, Cr\index{Cr} and Mn\index{Mn} in V\index{V} are
given in Fig. \ref{fig:3dsubsV}.
The calculated values are lower than the experimental values, although
the value for Cr\index{Cr} lies fairly close.
The Mn\index{Mn} resistivity is much larger than the other ones.
The value measured for the $5d$ impurity  Ta\index{Ta} of
$1.5 \mu\Omega{\rm cm/at\%}$ is very close to the calculated
value of $1.3 \mu\Omega{\rm cm/at\%}$.

\begin{figure}[b]
\centerline{\epsfig{figure=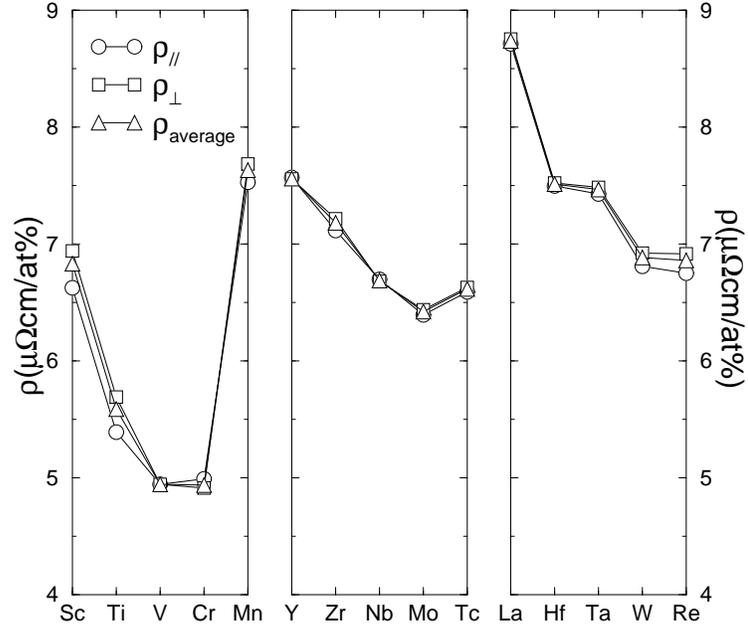,height=9.10cm}}
\caption{Calculated resistivities of $3d$, $4d$ and $5d$ impurities,
located
next to a vacancy
in V\index{V}.
\index{Sc}\index{Ti}\index{Cr}\index{Mn}
\index{Y}\index{Zr}\index{Mo}\index{Tc}\index{Nb}
\index{La}\index{Hf}\index{W}\index{Re}\index{Ta}}
\label{fig:rhovacV}
\end{figure}

The calculated resistivity of a vacancy in V\index{V} is larger than of
any of the $3d$ impurities\index{Sc}\index{Ti}\index{Cr}\index{Mn},
namely $4.94  \mu\Omega{\rm cm/at\%}$.
This results in resistivities of impurity-vacancy pairs,
varying from 5 to 9 $\mu\Omega{\rm cm/at\%}$,
as can be seen from Fig. \ref{fig:rhovacV}.
The large value for the Mn\index{Mn} impurity is also seen in the
$3d$ series in the left panel of the figure, but the effect is
\index{Sc}\index{Ti}\index{Cr}\index{Mn}
not as pronounced as in the case of a single impurity.
The resistivity turns out to be fairly isotropic, {\it i.e.}
$\rho_{\parallel}\approx\rho_{\perp}$ in Eq. \ref{eq:jBCC}.

It is seen that the resistivity of a $4d$
\index{Y}\index{Zr}\index{Mo}\index{Tc}\index{Nb}
impurity next to a vacancy
tends to be larger than that of a $3d$\index{Sc}\index{Ti}\index{Cr}\index{Mn}
impurity and smaller than that
of a $5d$
\index{La}\index{Hf}\index{W}\index{Re}\index{Ta}
impurity.
The resistivity for the $3d$ impurities is the lowest for V\index{V}, while for
the $4d$ impurities it is lowest for Mo\index{Mo}, which has an additional valence
electron.
For the $5d$
\index{La}\index{Hf}\index{W}\index{Re}\index{Ta}
impurities the resistivity of the impurity-vacancy pair
decreases monotonically with the atomic number.

\begin{figure}[t]
\centerline{\epsfig{figure=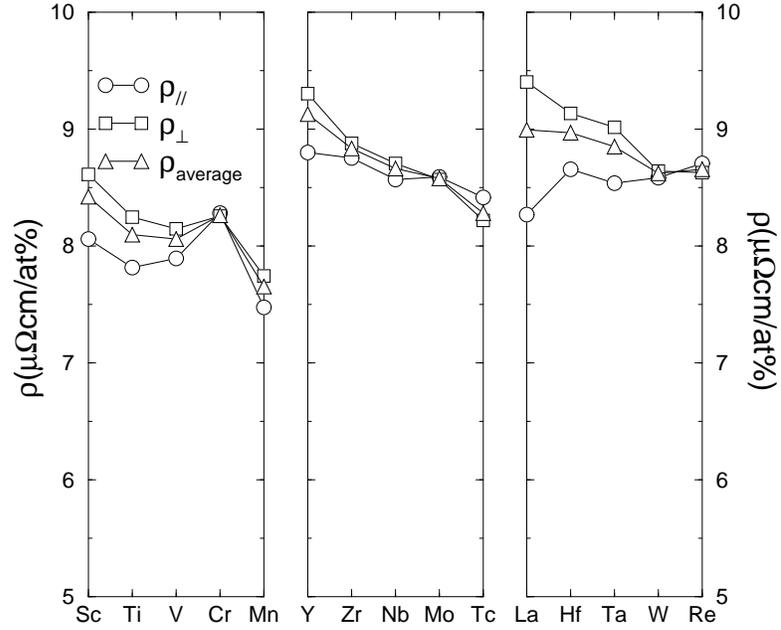,height=9.10cm}}
\caption{Calculated resistivities of $3d$, $4d$ and $5d$ impurities,
located
at the saddle point position
in V\index{V}.\index{Sc}\index{Ti}\index{Cr}\index{Mn}
\index{Y}\index{Zr}\index{Mo}\index{Tc}\index{Nb}
\index{La}\index{Hf}\index{W}\index{Re}\index{Ta}}
\label{fig:rhozptV}
\end{figure}

The resistivities for impurities at the saddle point are depicted in
Fig. \ref{fig:rhozptV}.
They show a larger anisotropy.
Exceptions are Cr\index{Cr}, Mo\index{Mo} and W\index{W}.
Apart from the high value of Cr\index{Cr}, the resistivity seems to decrease
monotonically in all three series.
The low value for Mn\index{Mn} is striking in view of the high values for the
single impurity and the impurity-vacancy pair.
The saddle point resistivities are larger than the initial point
values.
The small vacancies on either side of the atom could even enhance
this effect.

\section{Summary}
\label{sec:summaryres}

In this paper a multiple-scattering method has been described for
the calculation of the impurity resistivity.
It makes use of the calculated wave function coefficients, introduced
by Dekker {\it et al.} \cite{5:DekLodEk97FCC}.
The linearized Boltzmann equation
can be solved iteratively.
One iteration step involves the calculation of a Fermi surface
integral.
The integrand is the product of the vector mean free path, which
depends on the crystal momentum,
and two host wave function coefficients.
In its present formulation, the method is suitable to handle complicated
defects such as an atom during a diffusion jump.
It has been used to calculate the resistivity due to impurities,
vacancies and pair defects in Al\index{Al}, Ag\index{Ag} and V\index{V}.

The resistivities of $3d$ and $4sp$ impurities in Al\index{Al} have been
\index{Al}\index{Ca}\index{Sc}\index{Ti}\index{V}\index{Cr}\index{Mn}
\index{Fe}\index{Co}\index{Ni}\index{Cu}\index{Zn}\index{Ga}\index{Ge}
\index{As}
calculated, basically in order to see if the calculations make sense.
This series of impurities was investigated before by several
authors \cite{5:SchMro78,5:BoeLodMol83,5:PapStePap94} and experimental
values are available \cite{5:LanBor82}.
Their calculated resistivities turn out to depend strongly on the
atomic electronic configuration, which is used to construct the
crystal potential of the alloy.
This is especially important for transition metal impurities,
where {\it e.g.} the energies of $3d$ and $4s$ levels are almost
equal.
In this series it is seen that the resistivity decreases with
atomic number, when the impurity has two $4s$ electrons.
The shape of the experimentally observed peak is reproduced,
when the impurity carries one $4s$ electron.

Another consequence of the construction of the potentials,
the lack of charge neutrality, can be repaired by adding surface
charge to the atomic sphere of the impurity.
This procedure enlarges most calculated values and improves the
agreement with experiments.
Especially for transition metal atoms with many $d$ electrons,
and for the $4sp$\index{Cu}\index{Zn}\index{Ga}\index{Ge}
\index{As} impurities the agreement becomes very good.
Apparently that the calculation takes the essential features of the 
scattering process into account.
The strong dependence on the electronic configuration as well
as on the addition of surface charge make it interesting to
use self-consistent potentials in our calculation.

A vacancy plays an important role in the diffusion process.
Its calculated resistivity in Al\index{Al} of 0.6 $\mu\Omega{\rm cm/at\%}$
is much smaller than the experimentally
obtained value of 3 $\mu\Omega{\rm cm/at\%}$.
The resistivity of a host Al\index{Al} atom, halfway along its jump path
to a neighbouring vacant site, depends on the
direction of the electrical current and it is different from its value
for the atom at its initial position.
Both the direction and position dependence give rise to fluctuations
in the resistivity on a timescale of $10^{-13}~{\rm s}$.
The value of 0.41 $\mu\Omega{\rm cm/at\%}$, which is the average
over all current directions, is smaller than the value at the initial
position, the latter being equal to the resistivity of a vacancy.
In this calculation the two small moon-shaped vacancies next to
the jumping atom are not taken into account and it is expectable that
they will enlarge the resistivity.
The resistivity of a pair of vacancies depends on the direction
of the current.
If the pair is aligned with the current, the resistivity is smallest.
This can be attributed to a smaller cross-section for such a
configuration.
If the resistivity is averaged over all current directions it
equals the resistivity of two single ones.

The calculated resistivities due to the
$5sp$\index{Pd}\index{Cd}\index{In}\index{Sn}\index{Sb}
impurities in Ag\index{Ag} show a similar dependence on atomic
number as the experimental values. \cite{5:DekLodEk97FCCwind}
Just as for impurities in Al\index{Al} the resitivities are underestimated.
However, after achieving charge neutrality by adding a surface charge
to the impurity, they become too large.
The resistivity due to an impurity-vacancy pair is smaller than the sum
of the impurity and vacancy resistivities.
When the pair is aligned with the current, the resistivity is largest
and approximately equals that sum.
The fact that the resistivity is largest in that direction is in
contradiction with the smaller geometrical cross-section.
An impurity halfway its jump path has a larger resistivity than the
impurity-vacancy pair in spite of the neglected small vacancies.

The calculated resistivities of the impurities Cr\index{Cr} and Ta\index{Ta} in the
BCC transition metal V\index{V} agree fairly well with experiment, while
the one of Ti\index{Ti} is underestimated.
The values for a $d$ impurity-vacancy pair and an impurity halfway
its jump path are larger than the ones for a single impurity.

In conclusion, it has been shown that the resistivity due to low-symmetrical
defects can be calculated accurately.
The calculated impurity resistivities compare reasonably well with the
available experimental material.
They may even improve when self-consistent potentials for the alloy
are used.

\section*{acknowledgement}

This work was sponsored by the Stichting Nationale Computerfaciliteiten
(National Computing Facilities Foundation, NCF) for the use of
supercomputer facilities, with financial support from the Nederlandse
Organisatie voor Wetenschappelijk Onderzoek (Netherlands Organization
for Scientific Research, NWO).

The authors wish to acknowledge the contribution of Mr. P. J. Harte 
to a well-designed computer program for the calculation of the
impurity resistivity.


\begin{thebibliography}{10}
\bibitem{5:DekLodEk97FCCwind}
Calculations of the electromigration wind force for this series of
impurities in Ag were published previously in Ref.
10.
\end{thebibliography}


\end{document}